\documentclass[final]{aipproc}

\layoutstyle{6x9}

\begin{document}

\title{COSY--11 : How will we remember it ?}



\author{Colin Wilkin\thanks{Email: cw@hep.ucl.ac.uk}}{
  address={Physics and Astronomy Department, University College London, London, WC1E 6BT, UK}
}
\begin{abstract}
A personal selection is made of the highlights of the COSY--11
physics program undertaken at the COoler SYnchrotron of the
Forschungszentrum J\"ulich. This has been particularly rich in the
field of strange and non--strange meson production in
proton--proton and proton--deuteron collisions. The results are
considered in relation to experiments carried out at other
facilities and with respect to their impact on theory.
\end{abstract}

\maketitle

%%%%%%%%%%%%%%%%%%%%%%%%%%%%%%%%%%%%%%%%%%%%
%% MAINMATTER
%%%%%%%%%%%%%%%%%%%%%%%%%%%%%%%%%%%%%%%%%%%%

\section{Introduction}

Isaac Newton wrote in a famous letter to Robert Hooke: \textit{If
I have seen further, it is by standing upon the shoulders of
giants}. One of these \textit{giants} was, of course, Nicolaus
Copernicus and it is an honor to give this talk in this, his Alma
Mater. Innovations in Physics cannot be seen in isolation --- they
rely heavily on previous work, both theoretical and experimental.
Interested people have gone before and (hopefully) will come
after.

I have been asked, as an informed outsider, to try to answer the
question of how the COSY--11 program fitted into the development
of one specialized branch of Physics?

\section{The COSY--11 facility}

COSY--11 is a very \emph{simple} but effective facility, designed
to measure the production of meson(s) in nucleon--nucleon
collisions near threshold. \emph{i.e.}\ at an energy a little
above the minimum necessary for the process to happen. It uses the
specific characteristics of the COSY = COoler SYnchrotron at the
Forschungszentrum J\"ulich. COSY can accelerate protons up to
nearly 3\,GeV and store them at a fixed energy (coasting beam) for
tens of minutes. This is achieved by having the particles confined
in a 180\,m vacuum ring by a series of magnets. The magnetic force
is just sufficient to counteract the natural tendency for the
protons to go in a straight line.

\begin{figure}
  \includegraphics[height=.35\textheight]{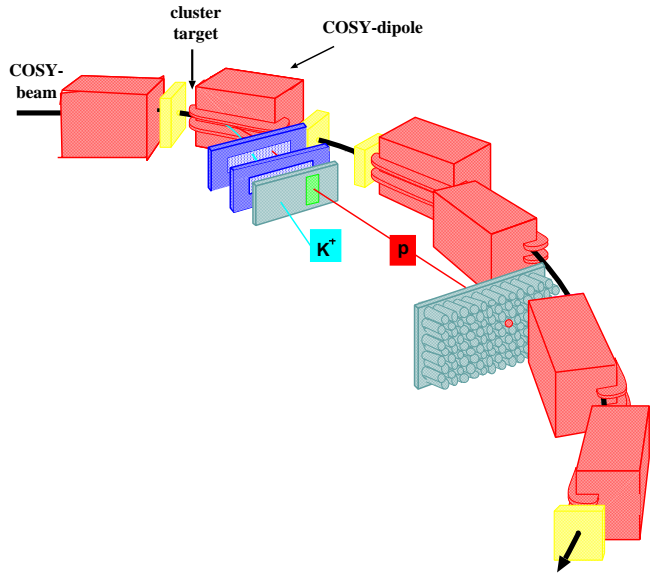}
\caption{Schematic diagram of the COSY--11 layout, as illustrated
in the COSY--11 calendar for January 1998.\label{Fig1}}
\end{figure}

The COSY--11 \emph{trick} relies on the fact that, if the proton
loses energy through an interaction in the target, the magnetic
force is too strong and the particle is bent to the inside of the
ring by an amount that depends upon its momentum. It (or another
particle) is then detected by drift chambers, silicon pad
detectors, and/or scintillation counters. The basic layout and the
principal components are shown in the schematic diagram of
Fig.~\ref{Fig1}, but more detailed descriptions are to be found
elsewhere in these proceedings.

Most of our body mass is composed of protons and the
proton--proton ($pp$) interaction is fundamental. However, above a
certain energy mesons can be produced in $pp$ collisions and so
the $pp$ force cannot be studied in isolation --- it is part of a
coupled system of nucleons and mesons. For this reason alone it is
clear that one must measure production reactions. Because of the
uncertainty principle, if mesons with masses of several hundreds
of MeV/c$^2$ are created, the reaction is going to depend upon the
proton--proton force at very short distances --- perhaps even
shorter than the size of the proton itself.

\section{The production of non--strange mesons in proton--proton collisions}

To understand the methodology of the COSY--11 technique, we first
have to discuss the principles behind a missing--mass measurement.
Consider the reaction $pp\to ppX$. If the momenta and hence the
energies of two outgoing protons $p_1$ and $p_2$ are measured, and
those of the beam proton $p_b$ and target $p_t$ are known, then
the momentum $\mathbf{p}_X$ \underline{and} energy $E_X$ of the
system $X$ are also both known. The mass $m_X$ of the produced
meson can then be calculated from $m_X^2 = E_X^2-\mathbf{p}_X^2$.
Thus at COSY--11 one never measured the neutral meson $X$ directly
but rather inferred that it had been produced by considering the
kinematics of the other particles in the final state.

Let us look first at the case of $X=\eta'$. The $\eta'$ is a heavy
brother of the pion, belonging to the same fundamental
pseudoscalar nonet. However, its mass is even a bit bigger than
that of the proton. In order to compare results for mesons with
widely different masses, we normally present data in terms of the
kinetic energy in the exit channel. Thus we define the excess
energy of the reaction as being $Q = W-2m_p-m_X$, where $W$ is the
total center--of--mass (c.m.) energy.

\begin{figure}
  \includegraphics[height=.40\textheight]{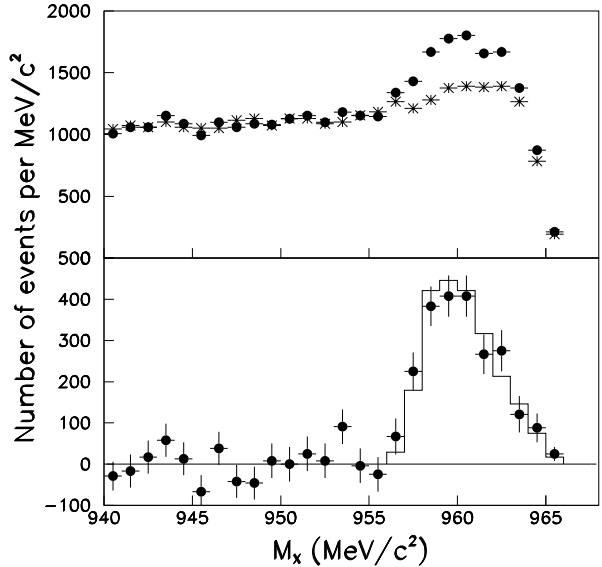}
\caption{Missing--mass spectrum of the $pp\to ppX$ reaction
measured with the SPESIII spectrometer at Saturne at a proton beam
energy of $T_p=2430\,$MeV~\cite{Hibou}. In the upper panel the
closed circles represent the actual data whereas the crosses are
estimates of the background obtained from measurements taken below
the reaction threshold. The differences are shown by the points in
the lower panel. These are compared to the Monte Carlo simulation
of the $pp\to pp\eta'$ reaction for $Q=8.3$\,MeV, which is shown
by the histogram.\label{Fig3}}
\end{figure}

The basic problems of a missing--mass experiment are the
minimization and evaluation of the background under the peak of
the $\eta'$ or other meson. The amount of background is usually
decided by doing measurements also below threshold. However, the
better the mass resolution, the easier it is to control the
background. Having subtracted the suitably scaled background from
the actual measurements, one ends up with a peak around the meson
mass. This is illustrated in the case of the SPESIII measurements
from Saturne~\cite{Hibou} which are shown in Fig.~\ref{Fig3}. The
next step is to make sure that the peak that has been generated in
this way has the right shape and this is done by doing a computer
(Monte Carlo) simulation, where one feeds in as much information
about the reaction and the experimental apparatus and conditions
as possible. For a spectrometer like COSY--11 or SPESIII, there is
a limited coverage of the angles, \emph{i.e.}\ there is a
restricted geometrical acceptance. Furthermore, in certain regions
the counters may have limitations and this reduces even further
the overall acceptance. The Monte Carlo simulations are of
tremendous importance in modern particle and nuclear physics
because they allow one to correct to some extent for these
deficiencies. The Monte Carlo of the $\eta'$ peak in
Fig.~\ref{Fig3} fits the data reasonably well and this gives some
confidence in the number of $pp\to pp\eta'$ events extracted.

\begin{figure}
 \includegraphics[height=.31\textheight]{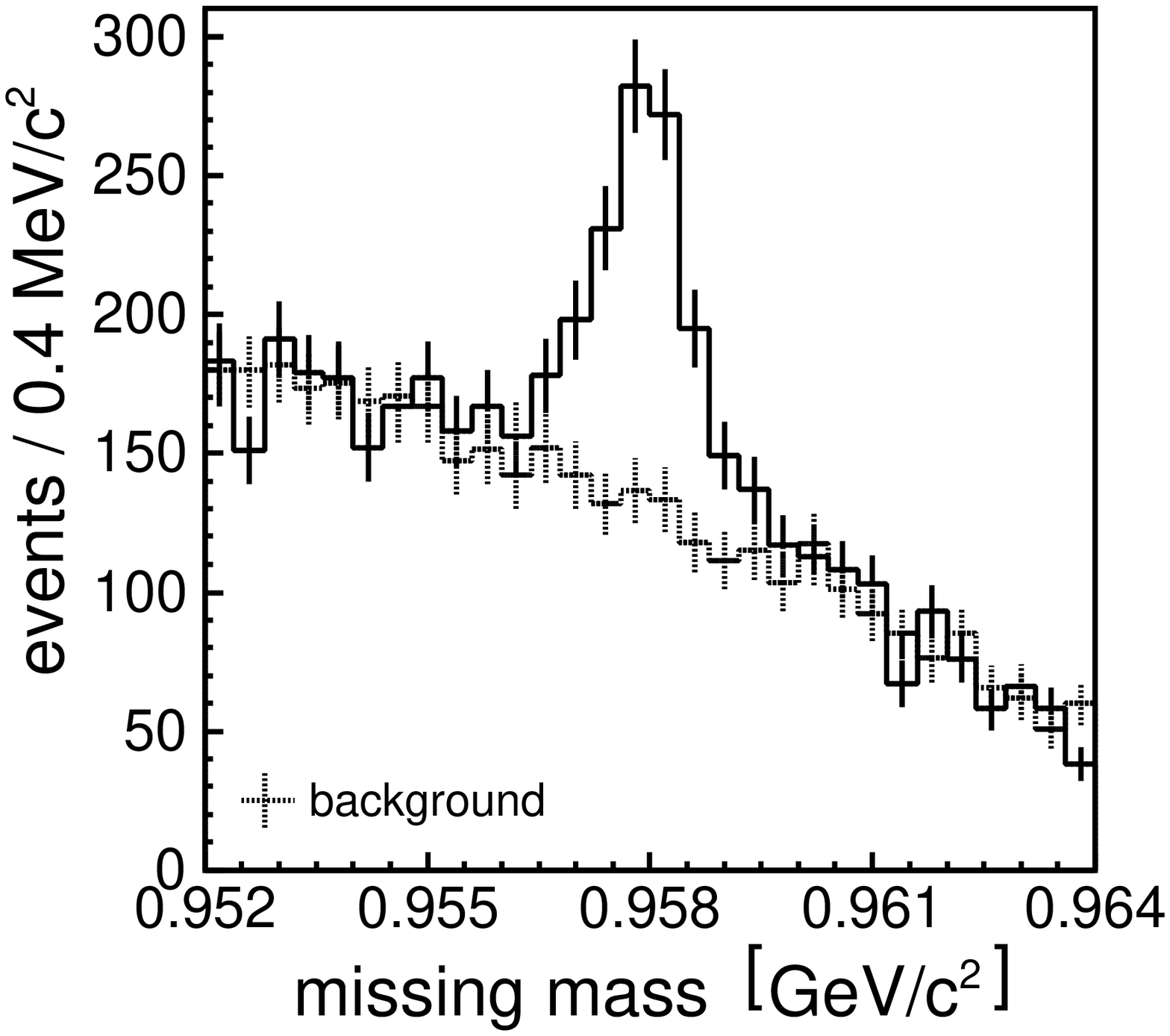}
 \includegraphics[height=.31\textheight]{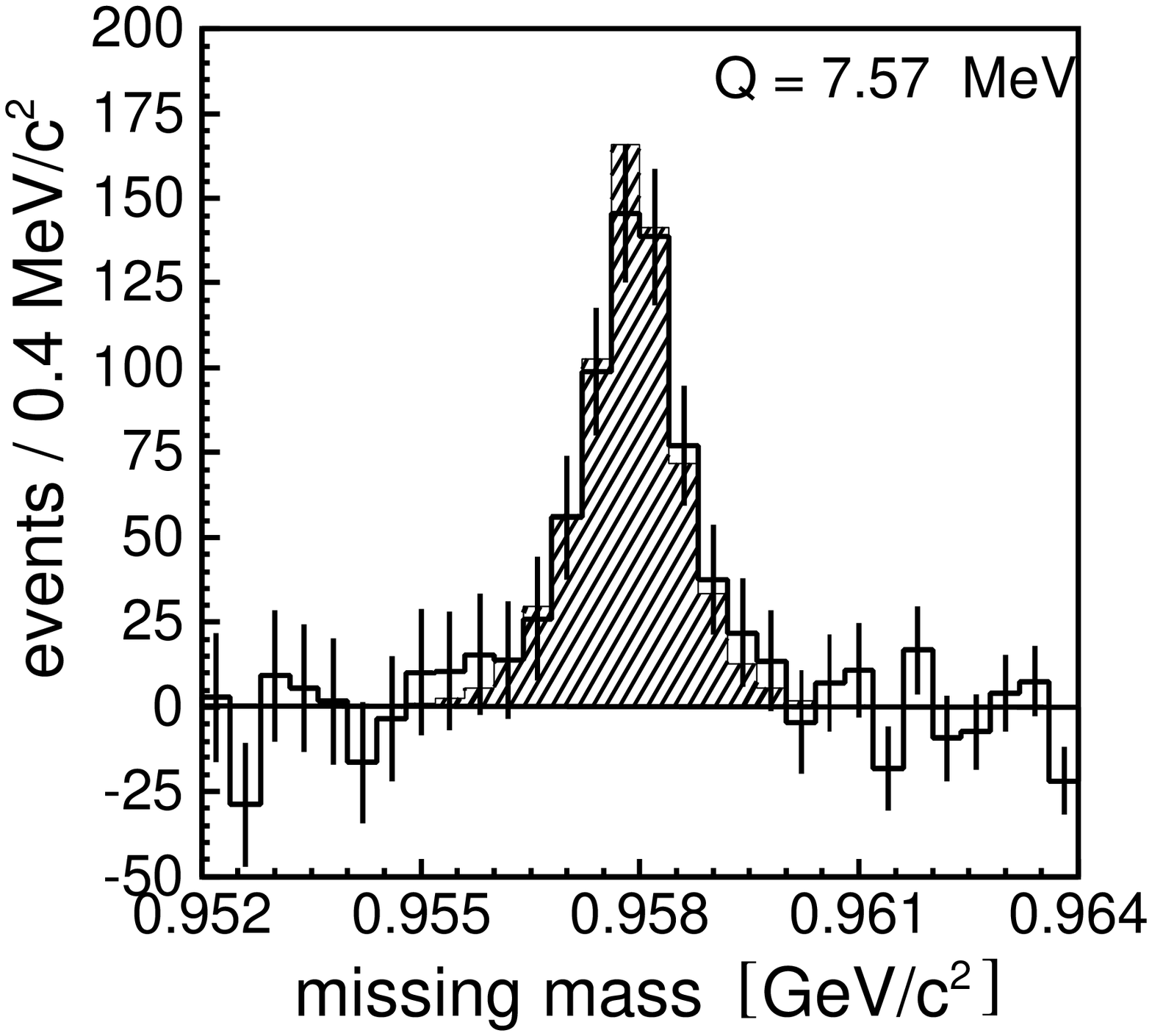}
\caption{Missing--mass spectrum from the $pp\to pp X$ reaction
obtained at an excitation energy of $Q=7.57$\,MeV with respect to
the $\eta'$ peak~\cite{Pawel2000}. The left panel shows the total
count rate in bins of 0.4\,MeV/c$^2$ (solid histogram) compared to
the background that is shown dashed. The difference in the right
panel is compared with the Monte Carlo simulation of $\eta'$
production. \label{Fig4}}
\end{figure}

The $pp\to pp\eta'$ reaction was studied at two energies at
SPESIII~\cite{Hibou} but it was measured at fifteen energies at
COSY--11~\cite{Pawel2000}, the results for one of which is shown
in Fig.~\ref{Fig4}. Both the excitation energy $Q$ and numbers of
$\eta'$ events were very similar to those of the SPESIII data of
Fig.~\ref{Fig3}. However, the better energy determination at COSY
allowed finer mass bins to be used (0.4 rather then
1.0\,MeV/c$^2$). This gives rise to a smoother background but it
is important to note that the shape of this background is
influenced by the specific characteristics of the spectrometer.
This is seen even more clearly for the $\eta'$ peak itself --- the
distribution is far more symmetric at COSY--11 and this is what
was to be expected on the basis of the Monte Carlo simulation that
is also shown in Fig.~\ref{Fig4}.

One very important parameter is the signal--to--background ratio
at the $\eta'$ peak. This is strongly influenced by the resolution
--- the narrower the peak the easier it is to isolate its
contribution to the production rate. For COSY--11 this is about
1:1 whereas at SPESIII it was only 1:3. It is clear that COSY--11
was very well suited for the study of such production experiments.

\begin{figure}
\includegraphics[height=.40\textheight]{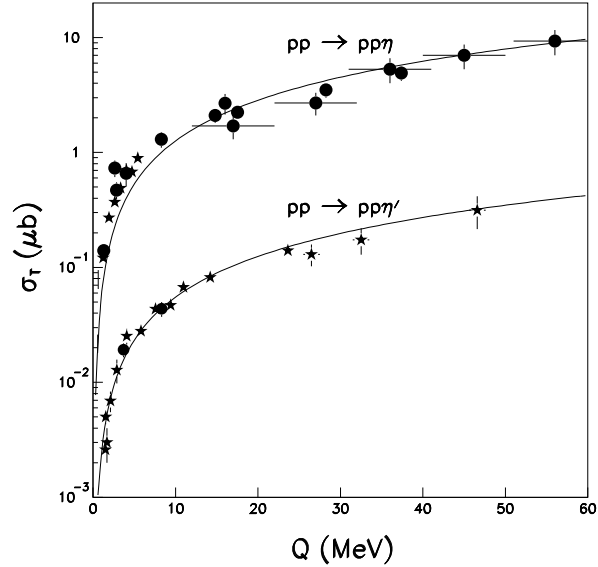}
\caption{ Variation of the total cross sections for $pp\to pp\eta$
and $pp\to pp\eta'$ with excess energy $Q$. The COSY--11
experimental results, shown by stars, are taken from
Refs.~\cite{Smyrski2000,Pawel2000,Pawel1998,Alfons2004}. Data of
other groups, shown by circles, are from
Refs.~\cite{Hibou,Calen1996,Calen1997}. The model of
Eq.~(\ref{formula}), that only includes phase space plus the $pp$
final state interaction~\cite{FW1996}, fails to describe the
$\eta$ production data in the near--threshold region, whereas it
is perfectly adequate for the $\eta'$. \label{Fig6}}
\end{figure}

The $pp\to pp\eta'$ total cross section varies smoothly with
excess energy Q when plotted on the logarithmic scale of
Fig.~\ref{Fig6}. Since the $\eta'$--nucleon force is believed to
be relatively weak, we would expect that this energy dependence
should be largely given by phase space, modified by the effect due
to the interaction of the two outgoing protons in the final state.
This is particularly strong in the $S$--wave because there is a
virtual state with a `binding' energy of $Q_0\approx 0.45\,$MeV.

If only the virtual bound state pole is taken into account, one
can easily obtain an analytic expression for the shape of the
energy dependence in the non--relativistic
approximation~\cite{FW1996}:
\begin{equation}
\label{formula} \sigma_T
=\sigma_0\frac{(Q/Q_0)^2}{\left(1+\sqrt{1+Q/Q_0}\right)^2}\,\cdot
\end{equation}
Although $Q_0$ is fixed just by the $pp$ interaction, the value of
$\sigma_0$ depends on the full dynamics of the $pp\to pp\eta'$
reaction. The $\eta'$ data of Fig.~\ref{Fig6}, which are
completely dominated by measurements from COSY--11, are well
represented by this description.

The $\eta$ meson has a much smaller mass than the $\eta'$ and
several groups have measured the total cross section for $pp\to
pp\eta$ and a selection of their results is shown in
Fig.~\ref{Fig6} along with those of $\eta'$ production. A
fascinating point seen here is that, unlike the case of $\eta'$
production, the \emph{ansatz} of Eq.~(\ref{formula}) cannot
reproduce simultaneously the low and higher $Q$ data. The curve
underpredicts the data below about 10\,MeV. This is almost
certainly the signal for a very strong interaction in the final
state, where the $\eta$ bounces between the two protons. For
heavier nuclei this final state interaction becomes so strong that
the $\eta$ becomes trapped for some time around the nucleus ---
but more of this later!

However, one also notices from Fig.~\ref{Fig6} that the cross
section for $\eta'$ production is about 25 times smaller than for
the $\eta$. It is a fairly general feature that, as the mass of
the meson is increased, the cross section drops. This is due
mainly to the momentum transfer from the initial to the final
protons rising with meson mass. On the other hand, the background
does not fall in the same way and it becomes increasingly
difficult to identify the meson purely through a missing--mass
peak. Coincidence measurements of the photons and pions arising
from the decay of the meson are then also required. This is one
part of the WASA program which has already started at
COSY~\cite{WASA}. For example, the reaction $pp\to pp\eta$ has
been identified cleanly through the decay chain $\eta\to
\pi^0\pi^0\pi^0 \to \gamma\gamma\gamma\gamma\gamma\gamma$. Thus,
although COSY--11 has been laid to rest, the succession is assured
through the WASA spectrometer~\cite{WASA}.

\section{Hyperon production in proton--proton collisions}

The second major COSY--11 success came from measuring $\Lambda$
and $\Sigma^0$ production in the $pp\to K^+\Lambda p$ and $pp\to
K^+\Sigma^0 p$ reactions. In such cases one has to study final
Kaon--proton correlated pairs instead of the proton--proton
required for $\eta$ and $\eta'$ production.

\begin{figure}[htb]
\includegraphics[height=.40\textheight]{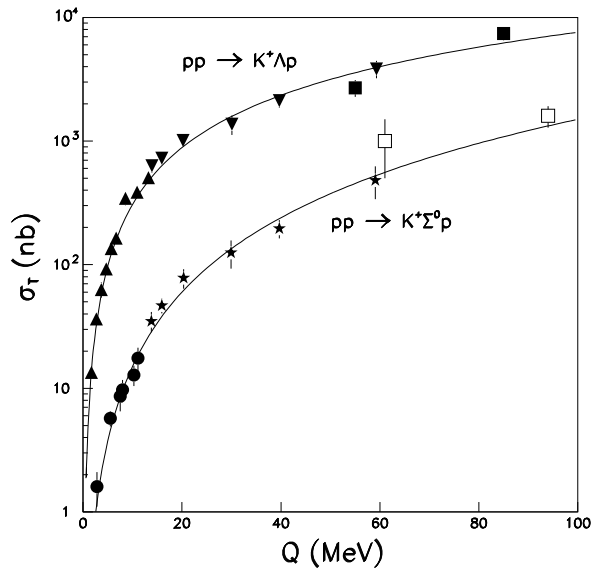}
\caption{Total cross sections for hyperon production in $pp$
collisions as functions of the excess energy $Q$. The experimental
points for the $pp\to K^+\Lambda p$ reaction come from COSY--11
(up and down triangles)~\cite{Balewski,Sewerin,Kowina} and
COSY--TOF (closed squares)~\cite{Bilger,Samad}. The curve
corresponds to the predictions of phase space moderated by a
final--state interaction between the $\Lambda$ and the proton. The
COSY--11 $pp\to K^+\Sigma^0 p$ points (closed circles and
stars)~\cite{Sewerin,Kowina} and two COSY--TOF points (open
squares)~\cite{Bilger,Fritsch} are compared to phase--space
predictions.
 \label{Fig7}}
\end{figure}

$K^+$ are much rarer in the final state than protons and it is not
possible to get an unambiguous identification just from the
available information on the trajectory and time of flight.
Therefore one takes the events around the kaon mass to see if the
missing mass then corresponds to a $\Lambda$ or a $\Sigma^0$. In
this way islands corresponding to the two reactions can be
isolated. Experiments could be done simultaneously at two
different energies to compensate for the mass difference of
$m_{\Sigma^0} - m_{\Lambda} = 77\,$MeV/c$^2$. For this purpose the
supercycle mode was used, where the beam was accelerated to one
energy and left to coast under stable conditions while the
$\Lambda$ production was measured. The beam energy was then raised
further to allow the $\Sigma^0$ to be studied before the energy
was lowered again. The results are shown in Fig.~\ref{Fig7}.

Apart from a couple of points from COSY--TOF~\cite{Bilger,Samad},
the $\Lambda$ data are dominated by results from
COSY--11~\cite{Balewski,Sewerin,Kowina}. Although the $K^+$
interacts weakly with the final proton, the $\Lambda p$
final--state interaction is known to be quite strong. The
$\Lambda$ does just bind with the deuteron and heavier
$\Lambda$--hypernuclei have been studied for fifty years. However,
though the $\Lambda p$ force is attractive, it is not quite
attractive enough to form a bound state and instead there is a
virtual state at an energy of $Q_0 \approx 5.5\,$MeV. If this were
the only distortion of phase space, the energy dependence of the
cross section would be given by Eq.~(\ref{formula}) and the
resulting curve does give a good description of the data in the
figure with this value of $Q_0$.

On the other hand, there is no convincing evidence for the
existence of $\Sigma$--hypernuclei. While this could be due to
their having very large widths engendered by the strong decay
$\Sigma N\to \Lambda N$, the COSY--11 data, which dominate
Fig.~\ref{Fig7}, offer a much simpler explanation. The $pp\to
K^+\Sigma^0 p$ total cross section follows a phase--space $Q^2$
variation, \emph{i.e.}, $Q_0\to\infty$ in Eq.~(\ref{formula}).
This is what we would expect if there were no strong interaction
between the $\Sigma^0$ and the proton. The obvious conclusion is
that there are no $\Sigma$--hypernuclei because the
$\Sigma$--nucleon force is not very attractive.

\begin{figure}[htb]
\includegraphics[height=.40\textheight]{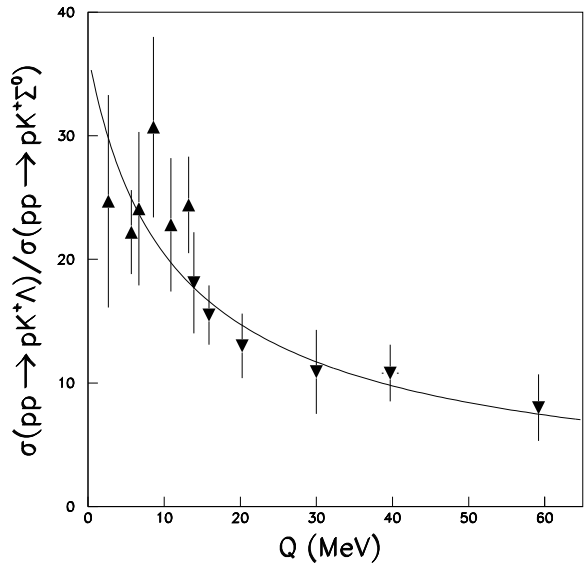}
\caption{The ratio of the total cross sections for the $pp\to
K^+\Lambda p$ and $pp\to K^+\Sigma^0 p$ reactions at the same
values of the excess energy $Q$. The COSY--11
points~\cite{Sewerin,Kowina} are compared to the predictions of
Eq.~(\ref{formula2}) where a final--state interaction is only
invoked in the $\Lambda p$ channel.
 \label{Fig8}}
\end{figure}

Because of the very different final--state interactions in the
$\Lambda p$ and $\Sigma^0p$ systems, the energy dependence of the
ratio of their production cross sections is very sharply peaked to
$Q=0$, as illustrated in Fig.~\ref{Fig8}. The data are well
described by just including the $\Lambda p$ interaction, and this
leads to:
\begin{equation}
\label{formula2}\sigma_T(pp\to K^+\Lambda p)/\sigma_T(pp\to
K^+\Sigma^0 p)
=R_0\left/\left(1+\sqrt{1+Q/Q_0}\,\right)^2\right.\,.
\end{equation}
Here $Q_0=5.5\,$MeV is the position of the virtual $\Lambda p$
state and $R_0\approx147$ depends upon the details of the dynamics
of both $\Lambda$ and $\Sigma^0$ production.

It is evident that COSY--11 was very well suited for the study of
hyperon production in proton--proton reactions. The obvious
questions left to be answered are:
\begin{itemize}
\item What is the production like in proton--neutron collisions?
\item What is the spin dependence of the production with polarized
beams and polarized targets?
\end{itemize}

Experiments at COSY--ANKE and/or COSY--TOF might help to resolve
some of these questions --- but they will not be easy to carry
out! Neutron detection have already been attempted at COSY--11 for
the $pn \to pn\eta'$ reaction using a deuterium
target~\cite{Joanna}. For this one has to measure both the low
energy \emph{spectator} proton from the deuteron target as well as
the fast final neutron. This is clearly an ambitious program given
the low overall acceptance but, even with only 19\% of the data
analyzed, a clear $\eta'$ signal is seen.

\section{Kaon--pair production in proton--proton collisions}

Two of the least understood of the light mesons are the scalar
(spin--0, positive parity) $a_0(I=1)/f_0(I=0)$ pair, both of which
have masses around 980\,MeV/c$^2$. They are found dominantly in
the $\pi\eta$ and $\pi\pi$ channels, respectively, but are also
`seen' in $K\bar{K}$~\cite{PDG}. Since the $K^+K^-$ threshold is
at 987.4\,MeV/c$^2$, this channel distorts the shapes of the
$a_0/f_0$ resonance peaks --- they may even be mainly molecular
rather than quark--antiquark states. The question asked at COSY
was whether one could measure the production of these resonances
in proton--proton collisions through the detection of their
$K^+K^-$ decay branch.

\begin{figure}[hbt]
\includegraphics[height=.40\textheight]{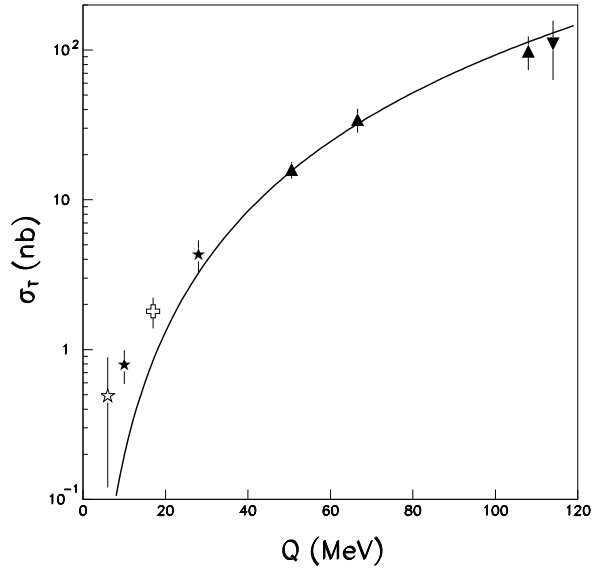}
\caption{Total cross section for the $pp\to ppK^+K^-$ reaction.
The experimental data are taken from Refs.~\cite{Wolke} (open
star), \cite{Quentmeier} (open cross), \cite{Winter} (closed
stars), \cite{Irakli} (triangles), and \cite{Balestra2001}
(inverted triangle). For data that were measured above the $\phi$
threshold, only the contribution from non--$\phi$ production is
reported here. The curve is the generalization of
Eq.~(\ref{formula}) to a four--body phase space, the analytic
formula for which is to be found in Ref.~\cite{Tengblad2006}.
 \label{Fig9}}
\end{figure}

The DISTO collaboration at Saturne measured the cross section for
$pp\to ppK^+K^-$ at quite high excess energy~\cite{Balestra2001}.
However, since the reaction involves a four--body final state, the
cross section drops very fast as threshold is approached. On the
other hand, the limited angular coverage meant that at COSY--11
the reaction could only be studied near threshold. The protons
were measured here as before, using the time of flight from a
start to a stop scintillator. Because of its smaller mass the
$K^+$ has a lower momentum near the reaction threshold and is
therefore bent more by the analyzing dipole. The start signal for
this particle was deduced from the measurement of the two protons
rather than being directly determined. One complication is that
kaons die quite quickly and this reduced significantly the overall
acceptance. Having identified the two protons, as well as a
candidate for the $K^+$, the final selection of the $pp\to
ppK^+K^-$ reaction was performed by evaluating the missing mass
with respect to the $ppK^+$ trio and making sure that it was equal
to that of the $K^-$.

\begin{figure}
\includegraphics[height=.40\textheight]{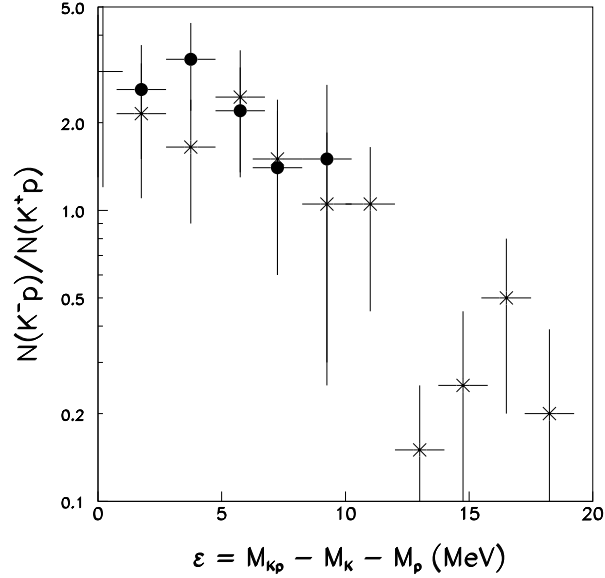}
\caption{Ratio of the number of events with the $K^-p$ and $K^+p$
in a given invariant mass bin measured from the $Kp$ threshold.
The COSY--11 data at $Q=10\,$MeV (closed circles) and 28\,MeV
(crosses) were reported in Ref.~\cite{Winter}. It should be noted
that the 10\,MeV results have been scaled up by a factor of about
3.5 so that they can be compared more easily with the higher
energy results. \label{Fig10}}
\end{figure}

The four lowest energy points in Fig.~\ref{Fig9} come from
COSY--11~\cite{Wolke,Quentmeier,Winter}, with the higher energy
data being measured at ANKE~\cite{Irakli} and
DISTO~\cite{Balestra2001}. The only distortion to the four--body
phase space considered in the curve is that due to the $pp$
final--state interaction~\cite{Tengblad2006} and, when this is
normalized to the high--$Q$ points, it undershoots considerably
the data at low $Q$. It therefore seems likely that there must be
an additional important final--state interaction among the
$ppK^+K^-$ particles. Direct evidence for this is to be found in
Fig.~\ref{Fig10}, where the ratio of the number of events
(corrected for acceptance) for $K^-p$ and $K^+p$ is plotted
against the invariant mass in the $Kp$ system. There are far more
events with low $K^-p$ invariant mass than $K^+p$. Since the
$K^+p$ force is known to be weak, the obvious conclusion to draw
is that there is also a strong $K^-p$ final--state interaction,
possibly connected with the influence of the $\Lambda(1405)$
hyperon resonance that is a little below the $K^-p$ threshold.

The COSY--11 measurements~\cite{Winter} were made below the
threshold for $\phi$ production and, for the analogous data from
ANKE~\cite{Hartmann,Maeda}, care has to be taken to distinguish
between $\phi$ and non--$\phi$ production. After this has been
done, the higher energy data show all the same features as those
presented in Fig.~\ref{Fig10} but with higher statistics. In fact,
the shape of the entire $N(K^-p)/N(K^+p)$ data set at different
energies can be described quantitatively through the introduction
of a universal $K^-p$ final--state interaction~\cite{Maeda}.

Analogous enhancements of the $\bar{K}^0d$ invariant--mass
spectrum are seen in the $pp\to d K^+\bar{K}^0$
reaction~\cite{Vera}. Thus the COSY--11 and the COSY--ANKE data,
taken together, quite clearly show that the $K^-$ is strongly
attracted to nucleons and this is likely to develop further for
heavier nuclei. On the other hand, the strength of the $K^-p$
force is such that it will be very hard to extract much
information on the $a_0/f_0$ complex from $pp\to ppK^+K^-$ or
similar reactions. However, physics is a continuous development.
The results in one experiment influence ideas at other facilities
and the scalar meson search will be continued at
COSY--WASA~\cite{WASA}.

\section{The $\mathbf{dp\to \,^3}\textrm{He}\,\eta$ reaction}

The last example that I have chosen to illustrate the COSY--11
legacy involves $\eta$ production in the three--nucleon sector.
When I discussed the $pp\to pp\eta$ reaction I stressed that,
unlike the case of the $\eta'$, the data seemed to require a
strong interaction of the $\eta$ with one or two of the final
protons.

\begin{figure}[htb]
\includegraphics[height=.40\textheight]{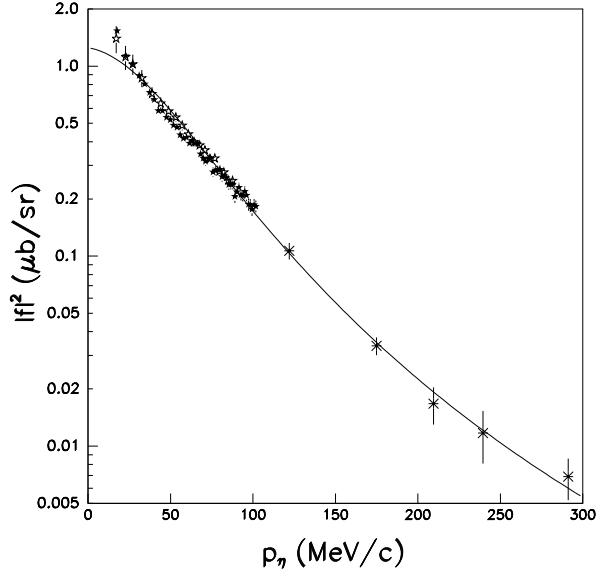}
\caption{Square of the backward amplitude for $dp\to
\,^{3}\textrm{He}\,\eta$. The refined COSY--11 (open
stars)~\cite{Smyrski2007} and COSY--ANKE data (closed
stars)~\cite{Timo2007} are hard to distinguish on this plot. The
very--close--to--threshold data, which are important in the
evaluation of the pole position, are not shown because of the
differences in the energy smearing in this region. The early
Saturne backwards' measurements are also shown, as is the
parameterization taken from the SPESIV paper~\cite{Berger1988}.
 \label{Fig11}}
\end{figure}

The situation is more extreme for $dp\to \,^{3}\textrm{He}\,\eta$
where it has even been speculated that the $\eta$ might even be
trapped, at least momentarily, while going around the $^3$He
nucleus~\cite{Wilkin1993}. Early Saturne measurements of the total
cross section near threshold~\cite{Berger1988,Mayer1996} gave
strong support for these ideas and the new very refined
measurements from COSY--11~\cite{Smyrski2007} and
COSY--ANKE~\cite{Timo2007} show conclusively that there is a pole
in the $dp\to \,^{3}\textrm{He}\,\eta$ scattering amplitude,
corresponding to a quasi--bound or virtual state, within about an
MeV of the threshold. The precise number depends upon exactly how
one takes the smearing of the beam momentum into account; this can
be important when one is discussing the difference between 1.0 or
0.5\,MeV!

The COSY--11 and COSY--ANKE experiments both benefited from the
negligible thickness of the target material and the ability of the
COSY machine crew to ramp up the energy in fine steps as a
function of time. Fortunately, the two data sets are completely
consistent within the systematic error bars, as is shown by the
plot of the square of the backward production amplitude shown in
Fig.~\ref{Fig11}. This geometry is chosen so that data at higher
energy~\cite{Berthet1985} could be shown on the same plot in order
to illustrate the continuity of the physics. The curve presented
in the figure is the parameterization of the SPESIV
low~\cite{Berger1988} and higher energy data~\cite{Berthet1985}.
However, the position of the quasi--bound (or virtual) state pole
in the complex $Q$--plane depends a lot on the data very close to
threshold and this is hard to obtain with an external beam and a
macroscopic target~\cite{Mayer1996}.

The energy dependence of the $dp\to \,^{3}\textrm{He}\,\eta$ total
cross section shows that the \underline{magnitude} of the
$s$--wave production amplitude $f_s$ decreases fast with excess
energy $Q$ (or the $\eta$ center--of--mass momentum $p_{\eta}$).
However, to prove conclusively that $f_s$ has a pole near $Q=0$ it
would be highly desirable to show that the phase of $f_s$ also
varies very rapidly in this region. Evidence for this comes from
the study of the angular dependence of the cross
section~\cite{Wilkin2007,Smyrski2007a}. For both the COSY--11 and
COSY--ANKE data it is seen that the differential cross section is
linear in the cosine of the $\eta$ angle with respect to the
initial proton:
\begin{equation}
\label{angle} \frac{d\sigma}{d\Omega}(dp\to
\,^{3}\textrm{He}\,\eta) \propto 1 + \alpha\cos\theta_{p\eta}\,.
\end{equation}

\begin{figure}[htb]
\includegraphics[height=.40\textheight]{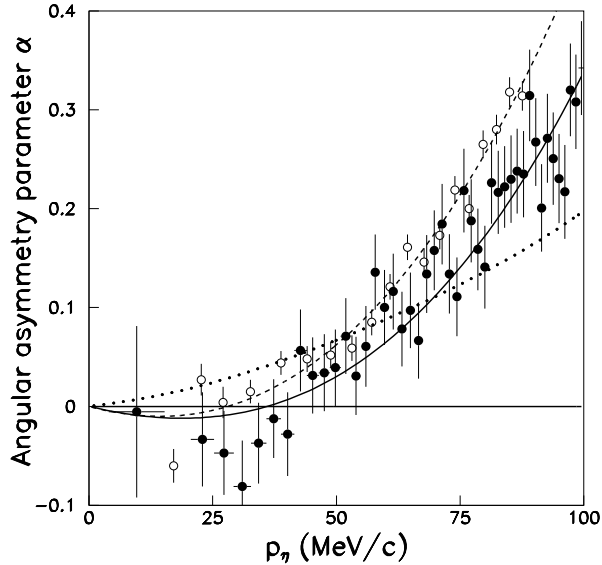}
\caption{Angular asymmetry parameter $\alpha$ of the $dp\to
\,^{3}\textrm{He}\,\eta$ reaction defined by Eq.~(\ref{angle}) as
a function of the $\eta$ c.m.\ momentum. The experimental data
from COSY--ANKE (closed circles)~\cite{Timo2007} and COSY--11
(open circles)~\cite{Smyrski2007} are compared to fits (solid and
broken lines) where the phase variation of the $s$--wave
production amplitude associated with the pole in the complex plane
close to $p_{\eta}=0$ is taken into account. If the phase
variation is neglected, the best fit (dotted curve) fails to
describe the data.
 \label{Fig12}}
\end{figure}

Near threshold, the asymmetry parameter $\alpha$ must arise from
an interference between $s$-- and $p$--waves of the final
$^{3}\textrm{He}\,\eta$ system and simple kinematic arguments
would suggest that $\alpha$ should vary linearly with
$p_{\eta}$~\cite{Wilkin2007}. This is, however, modified a little
because of the falling of the magnitude of $f_s$ with momentum.
Nevertheless, if one just takes into account the decrease of
$|f_s|$, one cannot reproduce the momentum dependence of $\alpha$
illustrated in Fig.~\ref{Fig12}. This parameter seems to be small,
or possibly even negative, for low $p_{\eta}$, and it is only
above about 40--50\,MeV/c that it starts to take off. This means
that the sign of the $s$--$p$ interference must be changing fast
with $p_{\eta}$ and hence that the phase of $f_s$ is a very
sensitive function of $p_{\eta}$. In Fig.~\ref{Fig12} are shown
the COSY--11 and COSY--ANKE data with crude fits to the $\alpha$
variation where the feedback of the $p$--waves to the total cross
section has been neglected. If, and only if, the phase variation
associated with the quasi--bound state pole is included in the
fits can the momentum variation of $\alpha$ be understood. More
precise modelling, where the total cross section and angular
distributions are fit
simultaneously~\cite{Wilkin2007,Smyrski2007a}, does not change
this conclusion in any material way. Both data sets require the
magnitude \underline{and} phase of the $s$--wave amplitude to vary
in the way expected if there is a pole in $f_s$ for small $|Q|$.

Production data of this kind can never prove whether the pole
corresponds to a quasi--bound or a virtual state but, in either
case, the COSY data have shown that there is a very unusual
nuclear system. This is a tribaryon with an excitation energy of
about 550\,MeV but a width of only a very few MeV. It is not, of
course, stable because it can decay with the emission of nucleons
and possibly pions. Hence one of the last of the COSY--11
experiments seems to confirm the existence of this very strange
nuclear state.

\section{Farewell COSY--11!}

I have lived through the closure of many machines --- the
Cosmotron at Brookhaven, the CERN SC, the Orsay SC, the Saturne
machine at Saclay, and the CELSIUS storage ring at Uppsala. What
really matters is the people and how they take what they have
learned at one facility and apply it in another environment.

Hadronic physics has advanced because of COSY--11 and there are
many possibilities to push even further forward, especially using
COSY--WASA~\cite{WASA}. The Polish--German collaboration has been
tremendously successful in personal terms; I have found them to be
a real delight to work with. The challenge is to ensure that this
atmosphere continues in the same way in the larger multinational
COSY--WASA setting.

\begin{figure}[hbt]
\includegraphics[height=.3\textheight]{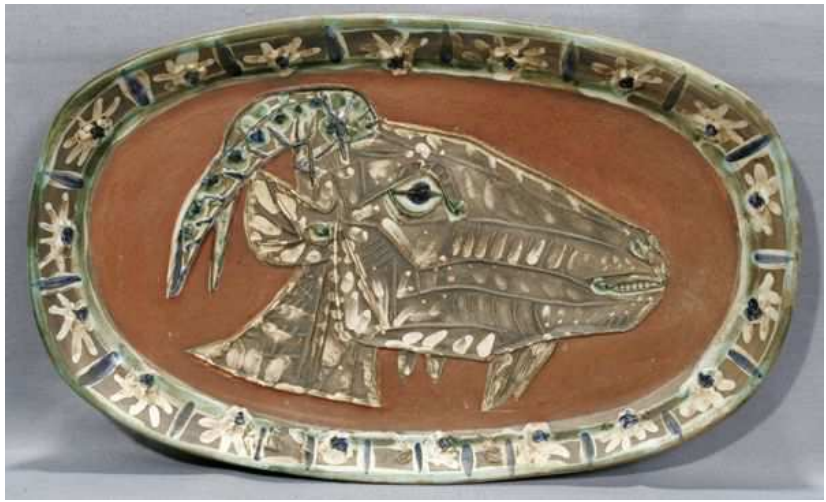}
\caption{Ceramic plaque of a goat stamped `Madoura Plein Feu' and
`Empreinte Originale de Picasso' on the Wilkin wall since 1967.
 \label{Fig13}}
\end{figure}

As has been made clear many times in this meeting, the symbol of
COSY--11 has been a goat. It is clear that I must have been a
supporter of COSY--11 long before the facility was conceived since
the ceramic plaque of a goat by Picasso shown in Fig.~\ref{Fig13}
has been hanging on a dining room wall in my house for the past
forty years. The message that I would like to leave you with is
that, to a first approximation, goats eat anything --- and
survive!

\begin{theacknowledgments}
I am very grateful to the organizers of the meeting for being
offered the opportunity to highlight some of the COSY--11
successes over the years. The selection has been very much a
personal one that has been tremendously biased by my own interests
and the presentation has been greatly influenced by my own views
on the physics involved. Finally, I hope that the goat will enjoy
its new pastures.
\end{theacknowledgments}
\bibliographystyle{aipproc}

\begin{thebibliography}{99}
%
\bibitem{Hibou} F.~Hibou \emph{et al.}, Phys.\ Lett.\ B \textbf{438}, 41--46
(1998).
%
\bibitem{Pawel2000} P.~Moskal \emph{et al}, Phys.\ Lett.\ B \textbf{474}, 416--422 (2000).
%
\bibitem{FW1996} G.~F\"{a}ldt and C.~Wilkin, Phys.\ Lett.\ B \textbf{382},
209--213 (1996).
%
\bibitem{Smyrski2000} J.~Smyrski \emph{et al.}, Phys.\ Lett.\ B \textbf{474},
182-187 (2000)
%
\bibitem{Pawel1998} P.~Moskal \emph{et al.}, Phys.\ Rev.\ Lett.\ \textbf{80},
3202-3205 (1998)
%
\bibitem{Alfons2004} A.~Khoukaz \emph{et al.},
Eur.\ Phys.\ J.\ A \textbf{20}, 345-350 (2004).
%
\bibitem{Calen1996}
H.~Cal\'en,\ \emph{et~al.,} Phys.\ Lett.\ B \textbf{366}, 39--43
(1996)
%
\bibitem{Calen1997}
H.~Cal\'en \emph{et~al.,} Phys.\ Rev.\ Lett.\ \textbf{79},
2642--2645 (1997)
%
\bibitem{WASA} \emph{WASA at COSY proposal}, Ed.\ B.~H\"oistad
and J.~Ritman, arXiv:nucl-ex/0411038.
%
\bibitem{Balewski} J.T.~Balewski \emph{et al}, Phys.\ Lett.\ B
\textbf{420}, 211--216 (1998).
%
\bibitem{Sewerin} S.~Sewerin \emph{et al.}, Phys.\ Rev.\ Lett.\ \textbf{83},
682--685 (1999).
%
\bibitem{Kowina} P.~Kowina \emph{et al.}, Eur.\ Phys.\ J.\ A \textbf{22},
293--299 (2004).
%
\bibitem{Bilger} R.~Bilger \emph{et al.}, Phys.\ Lett.\ \textbf{420}, 217--224 (1998).
%
\bibitem{Samad} S.~Abd El-Samad \emph{et al.}, Phys.\ Lett.\ B \textbf{632},
27--34 (2006).
%
\bibitem{Fritsch} M.~Fritsch, PhD thesis, University of
Erlangen--N\"urnberg (2002).
%
\bibitem{Joanna} J.~Przerwa, \emph{these proceedings}.
%
\bibitem{PDG} W.--M.~Yao \emph{et al.}, J.\ Phys.\ G \textbf{33}, 1--1232 (2006).
%
\bibitem{Balestra2001} F.~Balestra \emph{et al.}, Phys.\ Rev.\ C
63, 024004-1--15 (2001).
%
\bibitem{Wolke} M.~Wolke, PhD thesis, University of M\"unster
(1997).
%
\bibitem{Quentmeier} C.~Quentmeier \emph{et al.}, Phys.\ Lett.\ B \textbf{515},
276--282 (2001).
%
\bibitem{Winter} P.~Winter \emph{et al.}, Phys.\ Lett.\ B \textbf{635},
23--29 (2006).
%
\bibitem{Irakli} I.~Keshelashvili, PhD thesis, University of
Tbilisi (2006).
%
\bibitem{Tengblad2006} U.~Tengblad, G.~F\"{a}ldt, and C.~Wilkin,
Acta Physica Slovaca \textbf{56}, 205--211 (2006).
%
\bibitem{Hartmann} M.~Hartmann \emph{et al.}, Phys.\ Rev.\ Lett.\ 96,
242301-1--4 (2006).
%
\bibitem{Maeda} Y.~Maeda \emph{et al.}, (\emph{in preparation}).
%
\bibitem{Vera} V.~Kleber \emph{et al.}, Phys.\ Rev.\ Lett.\ \textbf{91},
172304-1--4 (2003);  A.~Dzyuba \emph{et al.}, Eur.\ Phys.\ J.\ A
\textbf{29}, 245--251 (2006); A.~Dzyuba \emph{et al.} (\emph{in
preparation}).
%
\bibitem{Wilkin1993} C.~Wilkin, Phys.\ Rev.\ C \textbf{47},
R938--940 (1993).
%
\bibitem{Berger1988} J.~Berger \textit{et al.}, Phys.\ Rev.\ Lett.\ \textbf{61},
919--922 (1988).
%
\bibitem{Mayer1996} B.~Mayer \textit{et al.}, Phys.\ Rev.\ C \textbf{53},
2068--2074 (1996).
%
\bibitem{Smyrski2007} J.~Smyrski \emph{et al.}, Phys.\ Lett.\ B \textbf{649}, 258--262 (2007).
%
\bibitem{Timo2007} T.~Mersmann \emph{et al.}, Phys.\ Rev.\ Lett.\ \textbf{98}, 242301-1--4
(2007).
%
\bibitem{Berthet1985} P.~Berthet \emph{et al.}, Nucl.\ Phys.\ A
\textbf{443}, 589--600 (1985).
%
\bibitem{Wilkin2007} C.~Wilkin \emph{et al.}, Phys.\ Lett.\ B (\textit{in press}),
arXiv:0707.1489.
%
\bibitem{Smyrski2007a} J.~Smyrski, \emph{these proceedings}.
%
\end{thebibliography}

\end{document}